# Title: Prefrontal scaling of reward prediction error readout gates reinforcement-derived adaptive behavior in primates

**Authors:** Tian Sang[1]†, Yichun Huang[1]†, Fangwei Zhong[2], Miao Wang[1], Shiqi Yu[1], Jiahui Li[1], Yuanjing Feng[3], Yizhou Wang[4], Sze Chai Kwok[5], Ravi S. Menon[6], Meiyun Wang[7*], Fang Fang[1*], Zheng Wang[1,8*]

**Affiliations:**

[1]School of Psychological and Cognitive Sciences; Beijing Key Laboratory of Behavior and Mental Health; National Key Laboratory of General Artificial Intelligence; IDG/McGovern Institute for Brain Research; Peking-Tsinghua Center for Life Sciences, Peking University, Beijing, China

[2]School of Artificial Intelligence, Beijing Normal University, Beijing, China

[3]Institute of Advanced Technology, College of Information Engineering, Zhejiang University of Technology, Hangzhou, China

[4]School of Computer Science; Center on Frontiers of Computing Studies; Institute for Artificial Intelligence; National Engineering Research Center of Visual Technology; National Key Lab of General Artificial Intelligence, Peking University, Beijing, China

[5]Phylo-Cognition Laboratory, DKU-The First People's Hospital of Kunshan Joint Brain Sciences Laboratory, Digital Innovation Research Center, Division of Natural and Applied Sciences, Duke Institute for Brain Sciences, Duke Kunshan University, Kunshan, China

[6]Department of Medical Biophysics, Centre for Functional and Metabolic Mapping, Western University, London, Ontario, Canada

[7]Department of Medical Imaging, Henan Provincial People's Hospital & the People's Hospital of Zhengzhou University, No. 7 Weiwu Road, Zhengzhou, Henan, China

[8]School of Biomedical Engineering, Hainan University, Haikou, Hainan, China

†These authors contributed equally to this work

*Correspondence to: Zheng Wang, PhD, Rm 336, LUI Che Woo Building, Peking University, Beijing 100871, China. E-mail: zheng.wang@pku.edu.cn

Fang Fang, PhD, Peking University, Beijing 100871, China. E-mail: ffang@pku.edu.cn

Meiyun Wang, MD, PhD, 7 Weiwu Rd, Henan Provincial People's Hospital, Zhengzhou, Henan, China. E-mail: mywang@ha.edu.cn

**Abstract:**

Reinforcement learning (RL) driven by reward prediction errors (RPEs) enables adaptive behavior across species. While midbrain-striatal encoding of RPEs is well-documented, how prefrontal circuits translate these signals into species-specific flexibility remains unknown. Integrating cross-species RL modeling, neuroimaging, and transcriptomics, we identified shared molecular and neuroanatomical infrastructure for RPE computation in humans and macaques. Despite this conserved foundation, humans demonstrated superior performance on deterministic and probabilistic reversal tasks. Single-trial decoding revealed equally robust dynamic RPE-to-choice prediction in both species; yet humans disproportionately recruited dorsal anterior cingulate (dACC) and dorsolateral prefrontal cortex (dlPFC). Interspecies alignment localized similarly predictive policy signals in macaque dACC/dlPFC that failed to guide behavior. Crucially, a within-species comparison in an independent human cohort demonstrated that only near-optimal performers exhibited strong prefrontal action decoding, contrasting with suboptimal individuals. These findings suggest that behavioral flexibility arises from prefrontal readout breadth - a hierarchical scaling principle that challenges efficient coding theories.

**Main text:** Prediction error - the mismatch between expected and achieved outcomes - is central to reinforcement learning (RL), serving as a teaching signal that drives adaptive decision-making[1-5]. Beyond midbrain dopaminergic circuits [6-10], reward-based trial-and-error learning by encoding reward prediction errors (RPEs) is now understood to involve extensive cortical integration, engaging the orbitofrontal (OFC) [11,12], ventral, lateral and dorsomedial prefrontal cortex [13,14], and dorsal anterior cingulate cortices (dACC) [15]. Together, these regions form a distributed cortico-basal ganglia network that translates RPEs into cognitive or behavioral adjustments [16-18]. Importantly, the presence of RPE signals does not necessarily guarantee their effective use in guiding behavior. Dysfunction in this system underpins maladaptive behaviors in psychiatric disorders [19,20], yet a crucial gap remains: while RPE-driven value updating is well documented, how RPEs shape behavioral performance [21-23], particularly the implementation of optimal choice strategies, remain unclear [24-26].

Reversal learning tasks, which require flexible strategy shifts under changing reward contingencies[27], offer an ideal setting to address these questions, as they elicit robust RPE while quantifying behavioral adaptation[28,29]. Within this task framework, humans and macaques share the capacity to update their choices based on recent outcomes as reward contingencies change, with comparative evidence suggesting broadly similar computational principles for integrating reward information[27,30]. Despite these commonalities, prior studies have also reported marked differences in reversal-learning performance in some task settings: humans can rapidly approach optimal behavior, whereas macaques may persist with suboptimal choices despite detectable RPE-related signals [14,29,31,32]. This apparent dissociation between RPE encoding and behavioral performance raises three critically overlooked questions about the biological processes converting RPE signals into observable behavior. First, the evolutionary conservation of RPE encoding mechanisms remains unclear, specifically whether computational principles, genetic substrates, and neuroanatomical implementations show cross-species preservation. Importantly, anatomical or functional correspondence alone does not establish whether RPE representations are supported by the same molecular substrates across species. Second, we lack mechanistic insight into how brain-wide, trial-to-trial fluctuating RPE signals are transformed into subsequent behavioral choices and whether this translation process itself undergoes species-specific adaptation. Third, and most paradoxically, species exhibiting comparable RPE computations demonstrate striking differences in behavioral adaptability, suggesting undiscovered mechanisms of this phylogenetically ancient signal beyond basic RL. We hypothesize that dlPFC and dACC, as key nodes known for conflict monitoring, rule maintenance and strategic updating [33-38], resolve this paradox by contextualizing RPE signals. These areas may dynamically gate how prediction errors throughout RL are interpreted and implemented by downstream circuits, potentially determining whether/how conserved RPE encoding yields species-divergent adaptive outcomes.

**Behavioral profiling reveals suboptimal and optimal strategies in reversal learning**

In a deterministic reversal-learning task (Fig. 1a), humans and macaques repeatedly chose between two options whose reward contingencies reversed periodically. The correct option was rewarded with 100% probability and the alternative with 0%; these contingencies were exchanged at each reversal. Thus, the first omission of reward from the previously correct option unambiguously signaled a reversal, and the optimal win–stay/lose–shift (WSLS) strategy maximized reward. To assess how closely each species followed this optimum under matched deterministic settings, we analyzed 32 sessions from 32 human participants and 207 sessions from three macaques.

Humans exhibited a higher proportion of WSLS-consistent reversals than macaques ($0.980 \pm 0.003$ versus $0.887 \pm 0.003$; Brunner–Munzel test, $P = 1.78 \times 10^{-113}$), switched more frequently after an unrewarded outcome ($0.962 \pm 0.009$ versus $0.707 \pm 0.008$; Brunner–Munzel test, Holm-adjusted $P = 2.25 \times 10^{-54}$), and made fewer additional errors after the first omission (0.017 ± 0.003 versus 0.105 ± 0.003; Brunner–Munzel test, Holm-adjusted $P = 2.09 \times 10^{-128}$; Fig. 1b–d). Reversal-aligned trajectories further highlighted this behavioral difference: humans typically committed a single informative error at reversal and recovered immediately, whereas macaques adjusted gradually over several subsequent trials. Quantifying deviation from the optimal WSLS trajectory via the full window mean L1 distance confirmed a smaller deviation in humans (0.007 ± 0.001 versus 0.134 ± 0.004; mean difference, -0.127; 10,000-iteration two-sided permutation test, $P = 1.00 \times 10^{-4}$; Fig. 1e–g). Consistent with these behavioral differences, model comparison favored StaticWSLS in humans but PosNeg-Inv-RW-p in macaques (Fig. 1h).

**Human behavior remains near-optimal under probabilistic feedback**

In a 75%/25% probabilistic reversal task (111 sessions from 37 participants; Fig. 2a), an unrewarded outcome could reflect either stochastic loss under the current contingency or a reversal, so adaptive adjustment required integrating consecutive outcomes rather than switching after every omission. Among fixed-threshold policies that switch after a given number of consecutive unrewarded trials, switching after two omissions yielded the highest mean reward rate (0.627 ± 0.002) and was used as the task-specific near-optimal strategy (Fig. 2b). Humans chose the currently high-reward option on 0.716 ± 0.005 of trials, close to but statistically lower than the 0.739 ± 0.005 achieved by the near-optimal strategy (mean difference 0.023; two-sided paired Wilcoxon signed-rank test, W = 1,522, P = 0.002; Fig. 2c).

Reversal-aligned analyses revealed temporally structured deviations: the near-optimal strategy outperformed humans at pre-reversal trials -2 (t(110) = -3.30, Holm-adjusted P = 0.008) and -1 (t(110) = -4.54, Holm-adjusted P = $1.00 \times 10^{-4}$), whereas humans outperformed the near-optimal strategy at post-reversal trial +2 (t(110) = 2.74, Holm-adjusted P = 0.036); no other time points differed after Holm correction (Fig. 2d). The mean L1 policy deviation from the near-optimal strategy in the probabilistic task was 0.084 ± 0.006. For comparison, the L1 deviation from the deterministic optimum was 0.013 ± 0.002 for humans and 0.152 ± 0.005 for macaques (Fig. 2e). Model comparison favored PosNeg-Inv-RW-p for human probabilistic behavior (model frequency = 0.913, exceedance probability = 1.000; Fig. 2f), with the corresponding group-level posterior parameter estimates and 95% credible intervals reported in Extended Data Table 5.

Taken together, these results show that humans remained near the task-specific near-optimal strategy under probabilistic feedback, but with reliable, temporally specific deviations around reversals. Across the two behavioral studies, humans closely followed optimal WSLS when a single omission fully identified a reversal and stayed near-optimal under outcome ambiguity, whereas macaques exhibited slower adjustment and greater deviation from the deterministic optimum. These behavioral differences motivated us to investigate whether suboptimal performance is reflected in the neural representation of RPE, which we next mapped in human and macaque task-fMRI datasets.

**Convergent neural representations of RPEs across species**

We next investigated whether the above RL computational principles extended to neural representations of RPEs. To enable valid cross-species comparisons, we implemented optimized fMRI protocols that accounted for each species' physiological constraints while ensuring equivalent data quality. Human participants performing the task underwent standard 3T scans, whereas awake macaques underwent functional imaging using customized motion-stabilizing helmets integrated with a customized four-channel surface receive coil (Fig. 3a; see Methods for details). Data quality from awake monkeys was rigorously validated through motion quantification and temporal signal-to-noise (tSNR) maps, revealing sub-voxel head motion (mean framewise distance = 0.15 ± 0.07 mm, median = 0.14, IQR = 0.09–0.18) and strong temporal stability (mean tSNR = 40.5 ± 24.1, median = 37.5, IQR = 21.9–56.3) (Fig. 3a, Extended Data Fig. S3a-d).

Using model-based fMRI with subject-specific RL regressors, we mapped whole-brain RPE-related activity in each species (Fig. 3c, FDR-corrected $P < 0.05$) based on the HCP_MMP1.0 parcellation (see Extended Data Table 4). Both humans and macaques exhibited significant RPE signals within a distributed cortico-striatal circuit, including the dACC (human: [8, 32, 22], t = 4.90; macaque: [-4.5, 9.5, 12], t = 2.63), dlPFC (human: [38, 50, 18], t = 6.01; macaque: [-18, 11, 10.5], t = 4.34), ventrolateral prefrontal cortex (vlPFC; human: [56, 26, 12], t = 4.29; macaque: [-21, 11, 0], t = 3.62), OFC (human: [22, 36, -16], t = 4.29; macaque: [7.5, 8, -1.5], t = 2.37), and insula (human: [32, 20, -8], t = 4.68). Subcortical regions also showed robust activation, including the caudate nucleus (human: [10, 0, 6], t = 5.56; macaque: [-4.5, 2, 3], t = 2.48) and putamen (human: [-18, 12, -10], t = 2.17; macaque: [-6, 5, 0], t = 2.67). Across the 21 sections defined in Extended Data Table 4, the human and macaque RPE maps showed high overall anatomical correspondence (Dice = 0.914; Jaccard = 0.842; detailed results are provided in Extended Data Table 6). These results delineate a conserved RPE network with broadly overlapping topographies in frontal and striatal areas.

**Transcriptional profiling of RPE representation networks across species**

Whether this neuroanatomical convergence also extends to the underlying molecular architecture, however, remains unknown, as an anatomically conserved trait, we examined transcriptional profiles of RPE representation networks using two species-specific transcriptomic atlases[39,40]. In macaques, we analyzed RNA-seq data from 38 bilateral regions using the D99 parcellation, while human analyses included 85 bilateral regions from the HCPex parcellation (see Methods). In macaques, differential gene expression analysis identified 2,483 genes significantly upregulated (FDR-corrected $P < 0.05$) in RPE-representing regions compared to the rest of the brain (Fig. 4a). KEGG pathway enrichment revealed pronounced involvement in neuromodulatory and synaptic signaling pathways, including: calcium signaling pathway (FDR-corrected $P = 1.58\times10^{-5}$, hypergeometric test), neuroactive ligand signaling ($P = 1.12\times10^{-9}$), oxytocin signaling pathway ($P = 4.95\times10^{-10}$), dopaminergic synapse ($P = 1.12\times10^{-7}$), glutamatergic synapse ($P = 6.13\times10^{-7}$), and cholinergic synapse ($P = 8.66\times10^{-6}$) (Fig. 4b). Gene Set Enrichment Analysis (GSEA) further corroborated these findings, highlighting significant enrichment (NES = 1.94, FDR-corrected $P = 0.03$) for response to dopamine (GO:1903350), driven by canonical dopaminergic genes (DRD1, DRD2, ADCY5, DRD3, and NSG2) (Fig. 4c).

In humans, 1,048 genes were significantly upregulated in RPE-representing regions (FDR-corrected $P < 0.05$, Fig. 4d). Mirroring macaque findings, KEGG enrichment highlighted dominant involvement of synaptic and neuromodulatory pathways, including: neuroactive ligand signaling (FDR-corrected $P$ = $1.58\times10^{-12}$, hypergeometric test), neuroactive ligand–receptor interaction ($P$ = $1.40\times10^{-4}$), dopaminergic synapse ($P$ = $4.15\times10^{-4}$), glutamatergic synapse ($P$ = $2.63\times10^{-4}$), serotonergic synapse ($P$ = $6.39\times10^{-4}$), and GABAergic synapse ($P$ = $4.15\times10^{-4}$) (Fig. 4e). GSEA reinforced this conserved theme, with significant enrichment for synaptic transmission, dopaminergic (GO:0001963; NES = 1.74; FDR-corrected $P$ = 0.04, Fig. 4f).

To identify evolutionarily conserved substrates, we mapped 10,252 orthologous genes between macaques and humans (HomoloGene Database). Among the upregulated orthologous genes (macaques: 1,644; humans: 728), 385 overlapped significantly ($P$ = $2.0\times10^{-128}$, hypergeometric test; Fig. 4g). Transcriptional changes were correlated across species (log2(fold-change): Pearson's r = 0.47, $P < 2.2\times10^{-16}$), indicating robust conservation. Protein–protein interaction (PPI) analysis (STRING) of the shared genes revealed five functional modules (Molecular Complex Detection algorithm, Fig. 4h, and Extended Data Fig. S4a): monoaminergic receptor signaling, cAMP regulation, synaptic organization, cell adhesion, and ephrin receptor signaling. Non-overlapping genes (343 human-only; 1,259 macaque-only; Fig. 4g) exhibited modest divergence (Extended Data Fig. S4b–c): human-specific genes enriched for receptor ligand activity (FDR-corrected $P$

= $2.16\times10^{-3}$, hypergeometric test) and Wnt signaling pathways ($P$ = 0.023); macaque-specific genes enriched for cytoplasmic translation ($P$ = 0.009) and ribosome ($P = 8.20\times10^{-5}$). Despite divergence, both species' conserved genes aligned with neuroactive ligand signaling ($P = 1.49\times10^{-5}$) and neuron to neuron synapse ($P = 2.94\times10^{-8}$). Interestingly, single-nucleus RNA-seq analysis revealed that shared upregulated genes were predominantly expressed in D1/D2 medium spiny neurons (Extended Data Fig. S4d-e), with secondary enrichment in astrocytes and interneurons. Together, these findings reveal substantial cross-species convergence in the molecular architecture of RPE-representing networks. Given this molecular and neuroanatomical conservation, the marked behavioral differences point to an unresolved source of cross-species divergence.

**Single-trial prediction of choice behavior via dynamic RPE signals**

We further investigated whether neural RPEs predicted choice behavior, potentially explaining species differences in behavioral optimality. By applying trial-wise fMRI analysis, we tracked how trial-by-trial RPE fluctuations govern subsequent choices and compared between species. We applied Least Squares–Separate modeling [17,41] to generate single-trial RPE activation patterns (one beta map per trial) from cortical parcels defined by group-level RPE maps (Fig. 5a). Using parcel-level RPE features, we trained a rigorously validated SVM classifier (100× nested cross-validation) to decode stay versus switch choices, achieving accuracies significantly above chance in both species (permutation test, $P < 0.001$; Fig. 5b, Extended Data Fig. S5). In macaques, decoding accuracy reached 63.8% (95% CI, 61.9–65.7%), reliably exceeding the null distribution (mean = 49.4% ± 4.1%; z = 3.49; Cohen's h = 0.28). Human performance was higher, with an accuracy of 69.0% (95% CI, 66.7–71.6%), also well above chance (null mean = 52.1% ± 3.2%; z = 5.25; Cohen's h = 0.39).

To identify brain regions driving cross-species behavioral prediction, we found that the OFC, vlPFC, dmPFC, frontal operculum, and subcortical structures (e.g., putamen, caudate) exhibited similar decoding contributions in both species (Fig. 5c). Despite statistically significant group differences (e.g., OFC: FDR-corrected $P < 0.001$; vlPFC: FDR-corrected $P < 0.001$), their species-specific contribution ratios remained near unity (OFC = 1.40; vlPFC = 0.74), suggesting conserved RPE encoding across primates. In contrast, human dACC and dlPFC showed markedly stronger decoding contributions than in macaques, with robust cross-species differences in effect size (dACC: t = –15.26, FDR-corrected $P = 2.9\times10^{-34}$, contribution ratio = 2.70; dlPFC: t= –65.53, FDR-corrected $P = 2.9\times10^{-135}$, ratio = 7.30).

**Human-aligned macaque RPE regions contain decodable information about best-policy actions**

We asked whether macaque regions functionally corresponding to the human RPE network contained information about actions prescribed by the best-performing policy, even when those actions were not consistently expressed in macaque behavior[42]. Using a cross-species joint-embedding framework (Fig. 6a), we projected human-defined RPE-predictive regions onto macaque cortex, identifying computationally similar areas independent of anatomy. We extracted trial-wise RPE signals from these aligned regions and trained an SVM classifier to predict task-optimal actions, with rigorous cross-validation as described above.

The classifier achieved 86.3% accuracy (95% CI: 78.9–89.9%, Fig. 6b), far surpassing chance (mean null = 62.7% ± 8.8%; $P < 0.001$; z = 2.67; Cohen's h = 0.81). Near reversal points, classifier predictions were significantly more accurate than observed monkey performance at multiple time lags (lag -3 ($t(128) = -4.63$, $P < 0.001$, 95% CI = [-0.135, -0.054]), lag -2 ($t(128) = -5.06$, $P < 0.001$, [-0.120, -0.053]), lag -1 ($t(128) = -5.22$, $P < 0.001$, [-0.123, -0.055]), lag 0 ($t(128) = 3.85$, $P < 0.001$, [0.023, 0.070]), lag 2 ($t(128) = -7.65$, $P < 0.001$, [-0.285, -0.168]), lag 3 ($t(128) = -3.17$, $P = 0.002$, [-0.149, -0.034]), except lag 1 ($P = 0.15$) (Fig. 6c). In a separate policy-distance analysis, cross-validated classifier predictions showed smaller L1 distances to the best-agent policy than did observed macaque behavior ($t(128) = -10.10$, $P < 0.001$), indicating that the decoded actions more closely approximated the optimal policy (Fig. 6d).

**Behavioral alignment with the benchmark policy predicts neural decoding of policy actions in humans**

Our cross-species analyses indicated that RPE-related neural responses persist even when behavior deviates from the optimal benchmark policy, yet the informational content of these neural signals for subsequent policy-consistent actions varies with behavioral performance. This study further investigated whether human RPE-related neural activity encodes optimal policy-defined actions in a unified task context, and whether encoding strength correlates with individual behavioral alignment to the optimal policy.

We analyzed an independent human behavioral-fMRI dataset acquired during a 75%/25% probabilistic reversal-learning task. We paired 114 behavioral records with participant-specific, history-matched action sequences generated by the optimal policy. The mean proportion of high-reward choices was $0.641 \pm 0.006$ for observed behavior and $0.657 \pm 0.005$ for policy-specified sequences (mean ± s.e.m.), with no significant difference between the two (two-sided paired Wilcoxon signed-rank test, W = 2662.5, P = 0.145; Fig. 7a).

Behavioral policy alignment was quantified for each participant as the proportion of valid trials in which individual choices matched optimal policy prescriptions. After stringent quality control, 85 participants were retained; excluding the median-ranked individual yielded equal-sized high- and low-alignment groups (n = 42 each). Across these 84 participants, policy alignment positively correlated with balanced accuracy for decoding optimal policy actions from prefrontal RPE-related patterns (Spearman's ρ = 0.491, 10,000-iteration permutation test, $P = 1.00 \times 10^{-4}$, Fig. 7b). Whole-brain FWE-corrected RPE maps revealed a larger extent of significant voxels in dlPFC for the higher-alignment group (8.92% vs. 2.75%), whereas dACC extent was similar (10.17% vs. 9.68%, Fig. 7c). Mean balanced accuracy was greater in the high-alignment group (0.628, 95% bootstrap CI [0.583, 0.669]) than in the low-alignment group (0.525 [0.501, 0.549]; group-label permutation test, $P = 2.00 \times 10^{-4}$, Fig. 7d), and both groups decoded above chance (high: $P = 1.00 \times 10^{-4}$; low: P = 0.044, two-sided sign-flip tests). These findings demonstrate that, within a single species and task, closer behavioral approximation to the optimal policy is associated with stronger prefrontal readout of policy-prescribed actions, arguing against confounds of species, paradigm, or training in the cross-species brain-behavior relationship.

## Discussion

Efforts to link RL models with whole-brain neural implementations across species have been hampered by methodological disparities, particularly the direct comparison of human fMRI with non-human primate electrophysiology [30,43-47]. Whether differences in adaptive performance reflect the representation of reinforcement signals or their translation into subsequent actions remains unresolved. We addressed this question using task-matched behavioral comparisons, computational modeling, cross-species fMRI, and an independent within-human analysis. Under matched deterministic 100%/0% contingencies, humans rapidly adopted the optimal win–stay/lose–shift strategy, whereas macaques adjusted more gradually and made additional errors (Fig. 1), consistent with previous observations[29]. Accordingly, human behavior was best captured by StaticWSLS, whereas macaque behavior was best captured by PosNeg-Inv-RW-p, consistent with prior evidence favoring feedback-dependent RL over WSLS in macaques[31]. In the human 75%/25% probabilistic task, where individual outcomes were ambiguous, behavior was also best captured by PosNeg-Inv-RW-p (Fig. 2). Thus, humans flexibly adapted their learning strategy to feedback informativeness, whereas macaques relied on incremental updating even under deterministic feedback. These findings identify the translation of RPE-related information into task-appropriate actions as a key candidate mechanism underlying variation in behavioral optimality. Cross-species fMRI nevertheless revealed highly overlapping RPE-related activity across distributed frontostriatal networks (Dice = 0.914; Jaccard = 0.842; Fig. 3), indicating that behavioral divergence coexisted with broadly conserved neural RPE representations.

We leveraged two unique transcriptome atlases of human and macaque brains to align with RPE representation networks for high-throughput screening molecular substrates predisposing both species to RL. Our functional network-based analyses identified for the first time a set of candidate genes relevant to RPE, revealing conserved upregulation of monoaminergic GPCRs, cAMP/$Ca^{2+}$ signaling effectors and synaptic regulators (e.g., DRD2, ADCY5, DRD3, RGS9) in RPE-representing regions. These components coalesce a canonical "GPCR→cAMP/$Ca^{2+}$→plasticity" axis - long hypothesized for RL [9,10,20] but now demonstrated transcriptomically across primates. Remarkably, complementary single-cell RNA-seq analysis confirms these signatures predominantly map onto D1/D2 medium spiny neurons, the principal integrators of dopaminergic RPEs (Extended Data Fig. S4). Beyond dopamine, serotonergic (e.g., HTR4) and GABAergic (e.g., GAD2) engagement observed in both species suggests broader neurotransmitter involvement in RPE than previously recognized [48]. Thus, conserved RPE computations rely not only on shared anatomy but also a deep molecular architecture optimized for neuromodulatory control and synaptic plasticity. Despite this overarching conservation, subtle species-specific divergence emerged in RPE transcriptional landscapes. Humans showed enrichment in receptor-ligand components (e.g., Wnt signaling), potentially enabling finer receptor regulation or developmental tuning of prefrontal circuits. Conversely, macaque-specific genes implicated ribosomal/translational pathways, suggesting heightened reliance on protein synthesis. Yet these differences were eclipsed by cross-species convergence in core monoaminergic and synaptic signaling (Fig. 4g). Collectively, these findings suggest that the marked behavioral divergence is less likely to reflect fundamentally different molecular substrates for RPE representation. Instead, they point to the possibility that species differences emerge more strongly in how this broadly conserved RPE system is read out to guide behavior.

How primate brains transform distributed RPE computation into flexible, adaptive decisions has long stood as a pivotal unresolved question in the field [17,25]. Our trial-by-trial neural-behavioral decoding showed that dynamic RPE signals within these co-activated regions reliably predicted next-trial choices in both species (Fig. 5). Strikingly, only humans disproportionately recruited dACC/ dlPFC carrying persistent cross-trial information, mechanistically connecting these regions to RPE-based decision optimization at single-trial resolution – an observation that refines debate surrounding conflict-control theory [33]. In macaques, RPE signals preceding optimal choices (e.g., lose-shift/win-stay) were weaker in dACC/dlPFC relative to humans (smaller activation amplitude and spatial extent), directly coupling RPE dynamics to future action selection [49]. This diminished dACC engagement plausibly restricts macaques' sensitivity to feedback, exacerbating reliance on past actions and impairing cross-trial adjustments. While dlPFC RPE signals were necessary for value-guided strategies, weaker recruitment led to unstable performance. Notably, individual macaques exhibited performance variability commensurate with dACC/dlPFC engagement. These findings suggest that while macaques and humans share core conflict-monitoring circuitry[32], quantitative differences in dACC/dlPFC implementation produce behavioral divergence.

The functional role of dACC/dlPFC in context-sensitive cognitive control is further supported by our interspecies alignment analysis: remapping human RPE signal coordinates to homologous macaque regions enabled accurate prediction of the behavior by the best-performing agent (Fig. 6). This is striking because macaques efficiently computed RPEs in prefrontal regions yet failed to translate them into optimal behavioral adjustments, a phenomenon we term the "Eternal Code". This interpretation is broadly consistent with findings in mouse frontal cortex showing that decision variables not currently guiding behavior can nevertheless be represented in population activity[50]. This unexpected encoding-decoding dissociation reveals a critical bottleneck: species- and individual-level performance differences likely originate not from cortical RPE encoding *per se*, but from downstream readout/decoding processes in dACC/dlPFC. This interpretation was further supported by an independent human dataset acquired during the same probabilistic reversal-learning task (Fig. 7). Participants whose behavior more closely approximated the benchmark policy showed stronger decoding of policy-consistent next actions from prefrontal RPE-related patterns. Higher-alignment participants also showed more extensive RPE-related engagement around dlPFC, whereas the corresponding extent around dACC was comparable between groups. Because species, task, reward contingencies, and training protocol were held constant within this analysis, these findings show that greater behavioral alignment with the benchmark policy was associated with stronger prefrontal readout of RPE-related information. Together with the cross-species results, they identify the capacity of distributed prefrontal circuits to transform conserved RPE signals into adaptive actions as a key neurocomputational candidate underlying variation in behavioral performance.

Intriguingly, our results establish that optimal subsequent choice prediction arises from broad-scale RPE signal pooling in primate dACC/dlPFC, revealing an integrative population coding scheme in frontal cortex during higher-order cognitive computation, a phenomenon we term “Prefrontal Scaling”. This represents a stark departure from the classic efficient coding theory governing in primary sensory areas [51,52]. Importantly, we first demonstrate that the disproportionate expansion of human prefrontal cortex provides a quantifiable neurocomputational edge in decision-making (~1.9-fold greater relative to macaques [53]). This evolutionary specialization emerged following the human-ape divergence ~5-7 million years ago and was subsequently amplified in Homo lineages. Furthermore, these results provide a parsimonious but unifying explanation for conflicting lesion studies [54], whereby incomplete dACC damage might spare behavioral flexibility to a certain degree despite impairing performance [33]. Nevertheless, this neural-behavioral dissociation mirrors psychiatric disorders where intact reward learning fails to guide adaptive choices, highlighting therapeutic potential in modulating action-selection mechanisms. While the fMRI study cannot disambiguate dACC/dlPFC sequential/concurrent dynamics (due to inherent limits of temporal resolution), our findings refine the longstanding evaluate-and-adjust model of prefrontal interactions [33,55-57]. By demonstrating that context-dependent choice adaptations emerge from selective recruitment of distributed RPE representations, not isolated regional activation, we extend current working models and identify a latent hierarchical scaling mechanism that could inform future development of RL-derived artificial intelligence, where scalability and flexibility are the ultimate benchmarks of success.

# Figure legends

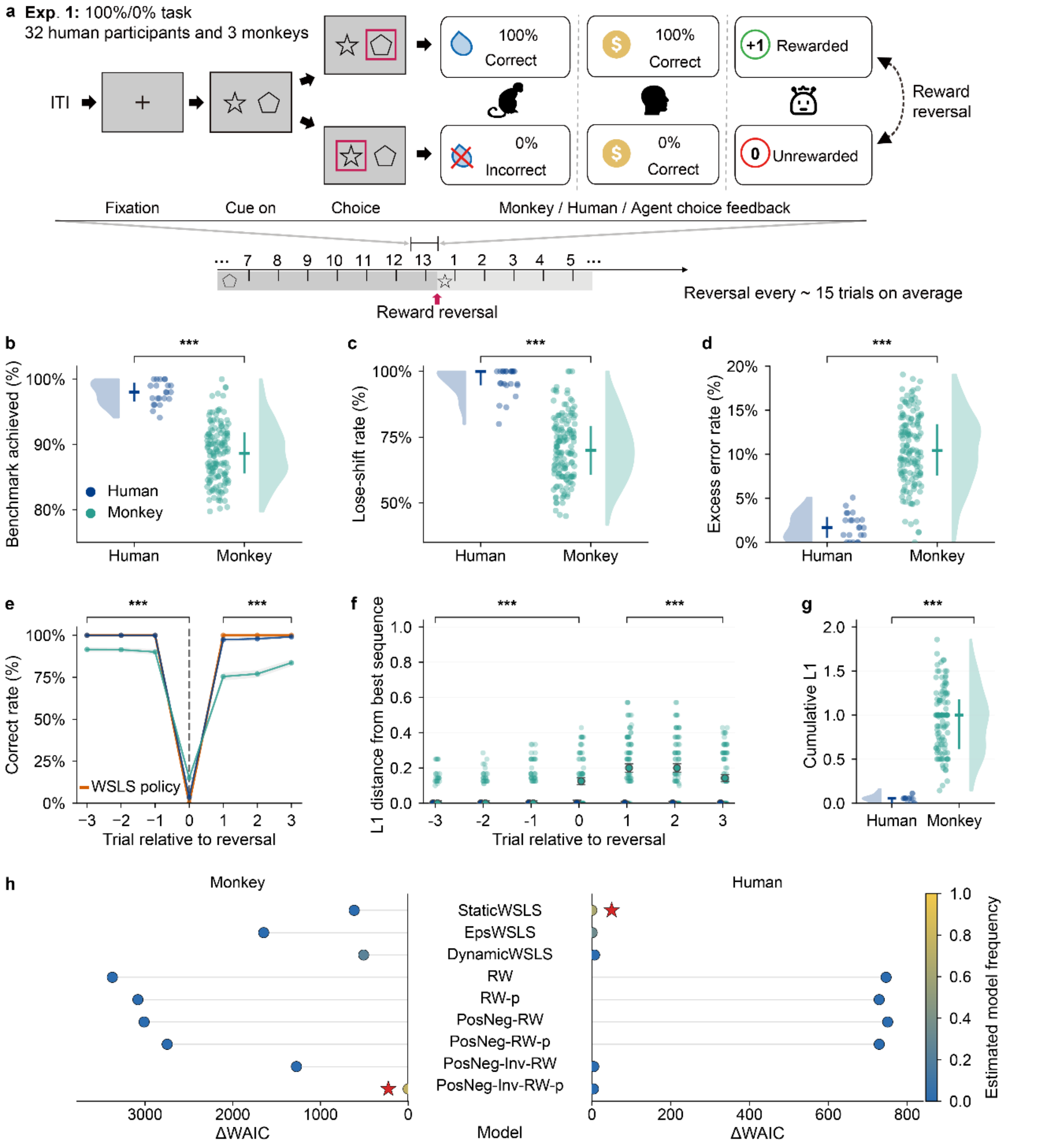

**Fig. 1: Deterministic reversal learning reveals species differences in adjustment after a single informative error.**

(a) Schematic of the deterministic two-choice reversal task. The correct option yielded reward with 100% probability and the incorrect option with 0%; these contingencies reversed without explicit cue. The first unrewarded choice after a reversal was designated the informative error.

(b) Proportion of reversals consistent with the optimal win–stay/lose–shift (WSLS) strategy. A value of 100% indicates no additional errors beyond the informative error. Humans showed a higher proportion than macaques (Brunner–Munzel test, $P = 1.78 \times 10^{-113}$).

(c) Lose-shift probability, the likelihood of switching after an unrewarded choice. Humans exhibited a higher lose-shift probability than macaques (Brunner–Munzel test, Holm-adjusted $P = 2.25 \times 10^{-54}$).

(d) Excess-error rate after the first informative error. Lower rates indicate fewer deviations from WSLS and faster adjustment. Excess-errors were less frequent in humans than in macaques (Brunner–Munzel test, Holm-adjusted P = 2.09 × 10-128). In panels (b–d), each point represents one human participant (n = 32) or one macaque session (n = 207).

(e) Reversal-aligned choice accuracy (probability of selecting the currently high-reward option). Trial 0 is the first trial after contingency reversal. The orange line denotes the optimal WSLS policy. Lines show group means, and shading indicates ±1.96 s.e. across pooled reversals.

(f) Reversal-aligned L1 distance between observed behavioral and the optimal WSLS policy; smaller values indicate closer agreement. Sessions with L1 values outside ±2 s.d. of the within-group mean at any displayed lag were excluded, retaining 21 human participants and 169 macaque sessions. Across the reversal window, the mean L1 distance was significantly lower in humans than in macaques (0.007 ± 0.001 versus 0.134 ± 0.004; two-sided 10,000-iteration permutation test, P = 1.00 × 10-4).

(g) Cumulative L1 distance across the reversal window for the same retained sessions as in f.

(h) Model comparison within the tested model set; lower ΔWAIC indicates better relative fit. The StaticWSLS had the highest estimated model frequency in humans (0.6400; exceedance probability = 0.9654), whereas the PosNeg-Inv-RW-p model had the highest estimated model frequency in macaques (0.7535; exceedance probability = 1.0000).

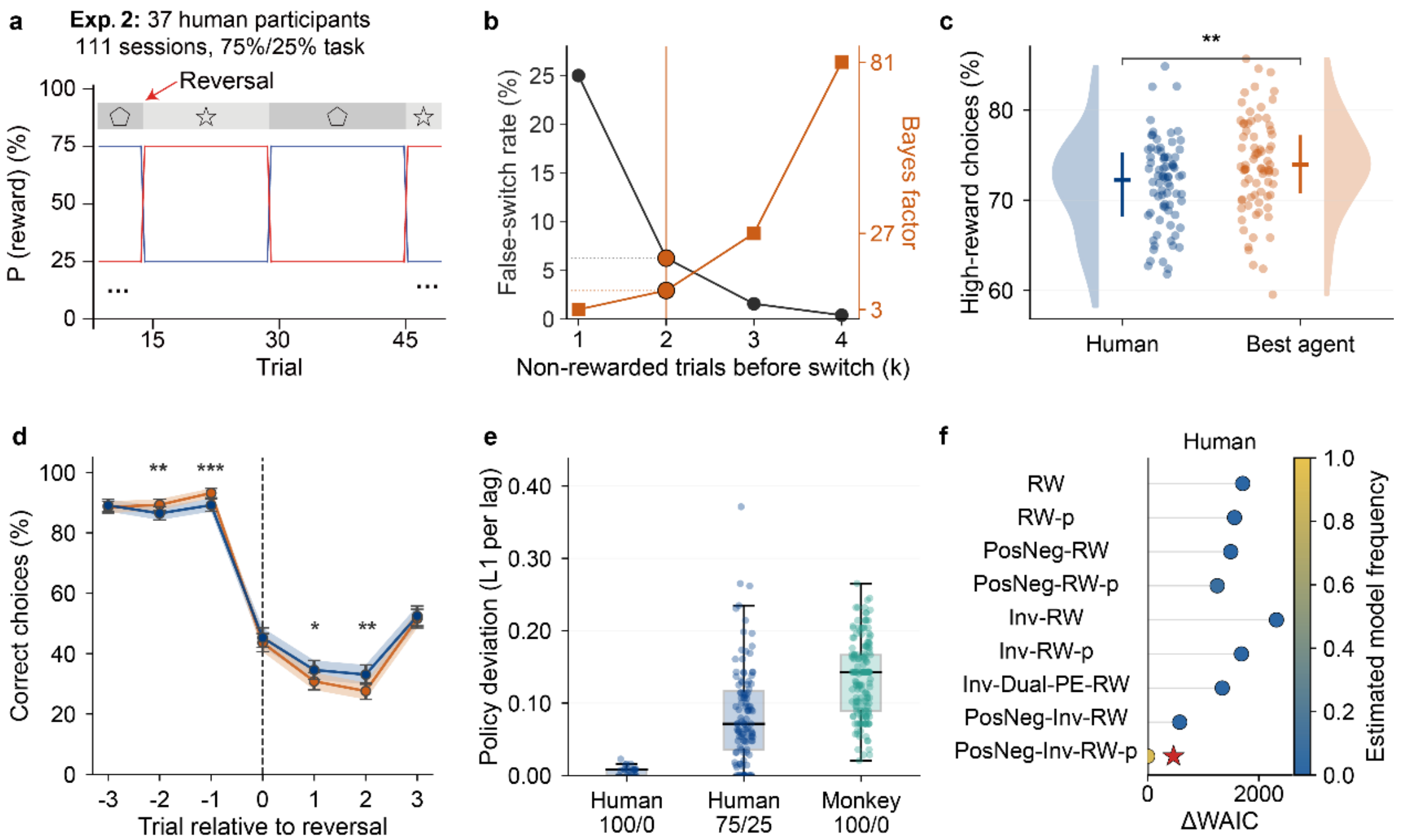

**Fig. 2: Human behavior approaches a task-specific near-optimal reference under probabilistic reversal learning.**

(a) Thirty-seven participants completed a probabilistic reversal-learning task (75%/25% reward contingencies, unsignaled reversals) across three runs of approximately 120 trials each.

(b) Fixed-threshold policies switching after k consecutive unrewarded choices (k = 1–4) were simulated under the task contingencies; k = 2 produced the highest mean reward rate (0.627 ± 0.002, 111 runs) and was used as the near-optimal reference, balancing rapid reversal detection against unnecessary switching (see Methods).

(c) Proportions of high-reward choices: human, 0.716 ± 0.005; best policy, 0.739 ± 0.005; two-sided paired Wilcoxon signed-rank test, W = 1522, P = 0.002).

(d) Reversal-aligned accuracy (mean ± 1.96 s.e.m.). The reference outperformed humans at trials -2 and -1 (Holm-adjusted P = 0.008 and <0.001), whereas humans outperformed the reference at trial +2 (P = 0.036); no other lags differed significantly.

(e) Mean L1 policy deviation across reversal lags: humans 100%/0% task, 0.013 (n = 32), humans 75%/25% task, 0.084 (n = 111); macaques 100%/0% task, 0.152 (n = 207 sessions). Lower values indicate closer alignment with the reference. All eligible sessions included, without L1-based outlier exclusion. (f) Model comparison for human probabilistic behavior: PosNeg-Inv-RW-p showed the highest model frequency (0.913, exceedance probability = 1.000), confirming better fit by a reinforcement-learning model than by WSLS-like alternatives, consistent with macaque results (Fig. 1h).

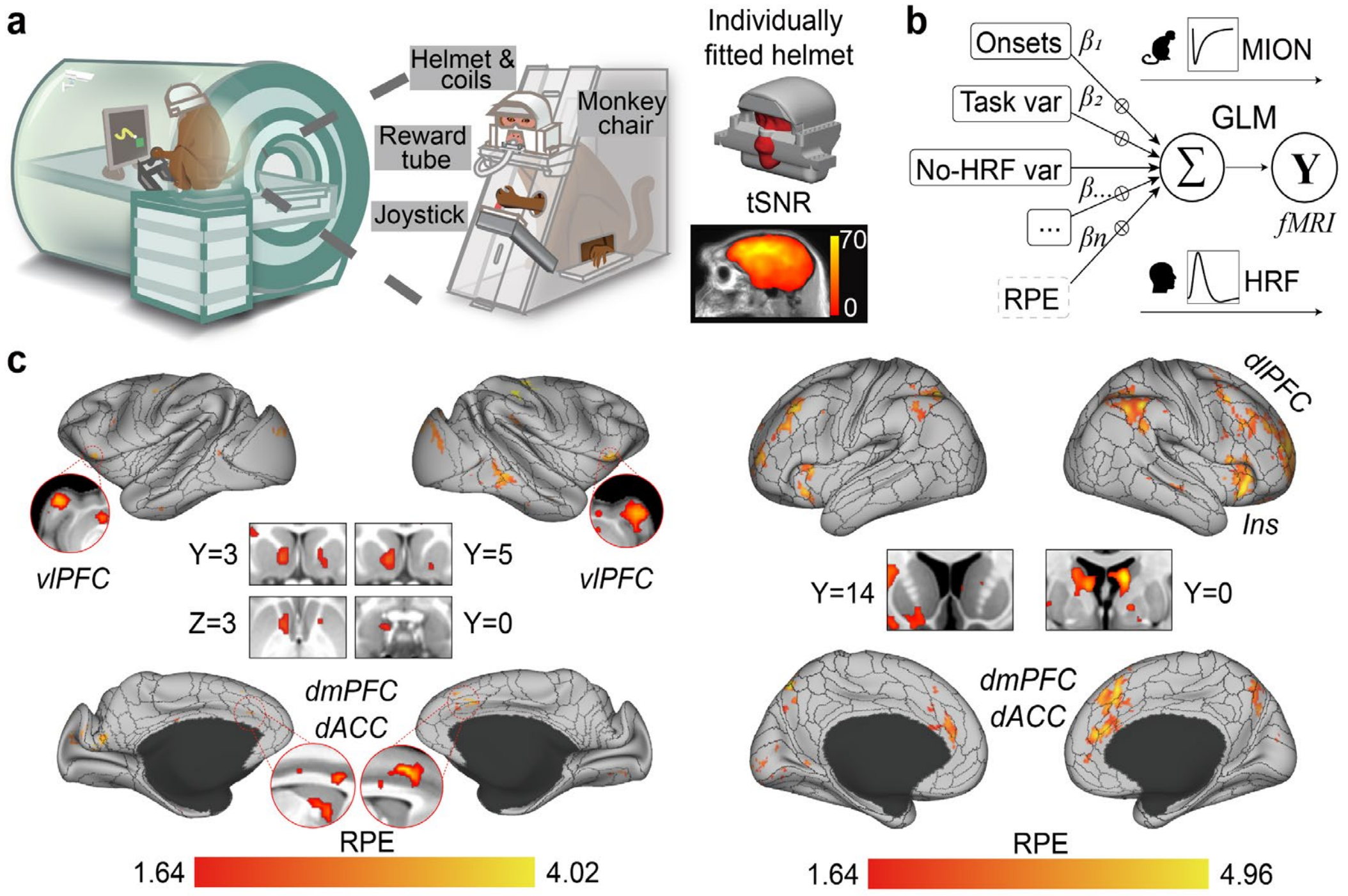

**Fig. 3: Neuroanatomically conserved representation networks of RPE across species.**

(a) Customized setup for awake macaque fMRI. Left: 3D-printed helmets (derived from individual skull scans) enabled non-invasive head fixation while maintaining natural posture. Right: Group-averaged temporal SNR (individual data in Extended Data Fig. S3d) confirmed protocol validity in awake macaques.

(b) RL Model-based fMRI pipeline for estimating RPE signals. Event-related regressors including onsets and parametric modulators were convolved with species-specific hemodynamic response functions (see Methods for details). Voxel-wise RPE effects were estimated using a general linear model (GLM).

(c) fMRI activation maps for RPE in macaques and humans. Voxel-wise GLM results were projected onto species-specific cortical surfaces for visualization. All maps were thresholded at $P < 0.05$, corrected for multiple comparisons using whole-brain FDR. Group-level consensus results for macaques are shown here (individual data in Extended Data Fig. S3e).

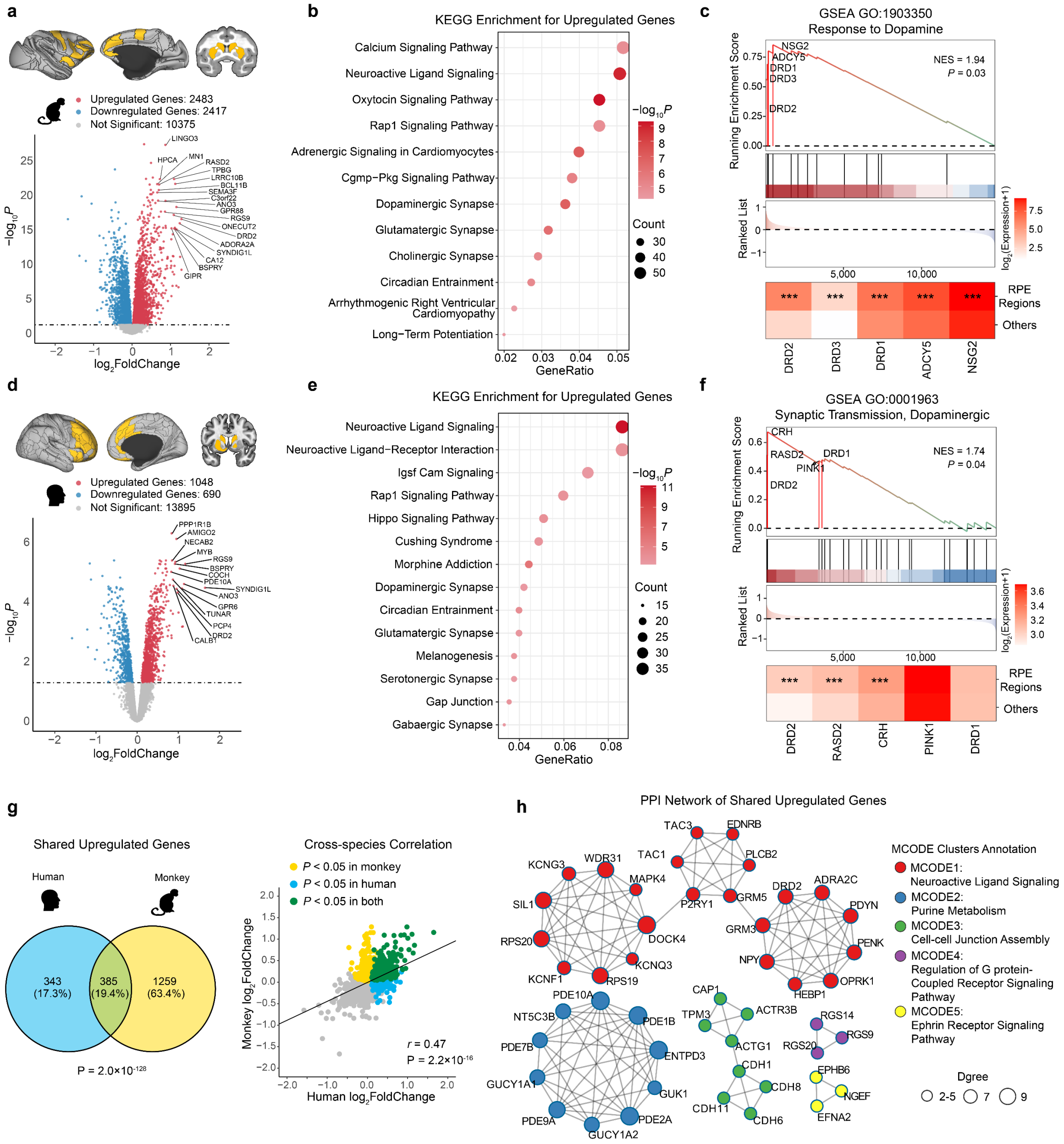

**Fig. 4: Molecular convergence in cross-species RPE representation networks.**

(a) Macaque RPE network transcriptomics. Volcano plot identifies 2,483 upregulated (red) and 2,417 downregulated (blue) differentially expressed genes (DEGs) versus non-RPE regions (FDR-corrected $P < 0.05$). Insets show representative macaque RPE regions. The dotted line represents FDR-corrected $P = 0.05$.

(b) Bubble chart represents the KEGG enrichment for upregulated DEGs in macaque. Top 12 significant terms are shown (FDR-corrected $P < 0.05$, hypergeometric test). The bubble size represents the number of overlapped genes between the gene list and each KEGG term and the color bar indicates the *P* value.

(c) The GSEA results of GO term "response to dopamine" (GO:1903350, NES = 1.94, FDR-corrected $P = 0.03$, GSEA permutation test) in macaque DEGs. The top five contributing genes are annotated for this term and their relative expression level are shown in the accompanying heatmap across RPE versus non-RPE regions (*** denotes FDR corrected $P < 0.001$).

(d) Human RPE network transcriptomics. Volcano plot identifies 1,048 upregulated (red) and 690 downregulated DEGs versus non-RPE regions (FDR-corrected $P < 0.05$). Insets depict representative human RPE regions. The dotted line represents FDR corrected $P = 0.05$.

(e) Bubble chart represents the KEGG enrichment for upregulated DEGs in human. Top 12 significant terms are shown (FDR-corrected $P < 0.05$, hypergeometric test). The bubble size represents the number of overlapped genes between the gene list and each KEGG term and the color bar indicates the *P* value.

(f) The GSEA results of GO term "synaptic transmission, dopaminergic" (GO:0001963, NES = 1.74, FDR-corrected $P = 0.04$, GSEA permutation test) in human DEGs. The top five contributing genes are annotated for this term and their relative expression level are shown in the accompanying heatmap across RPE versus non-RPE regions (*** denotes FDR corrected $P < 0.001$).

(g) Venn diagrams showing the overlap of upregulated homologous genes across species (left panel). Among these homologous genes, 385 (19.4%) were shared in both macaque and human ($P = 2.0 \times 10^{-128}$, hypergeometric test), with strong expression coupling (Pearson's $r = 0.47$, $P < 2.2 \times 10^{-16}$). The scatterplot distinguishes species-specific (yellow: macaque-only; blue: human-only) and conserved (green: both species) gene upregulation patterns.

(h) Representative protein-protein interaction (PPI) modules identified from shared upregulated genes (n = 385). Nodes represent proteins encoded by shared upregulated genes, and edges indicate known physical interactions (STRING interaction score > 0.4). Subnetworks were extracted using Molecular Complex Detection (MCODE) algorithm (minimum module size = 3 proteins). Five highest-scoring MCODE components are displayed. For each component, enrichment analysis was automatically applied to its constituent proteins, and the top enriched terms were used to annotate its functional category. Distinct colors denote MCODE-defined clusters, and node size corresponds to its connectivity degree.

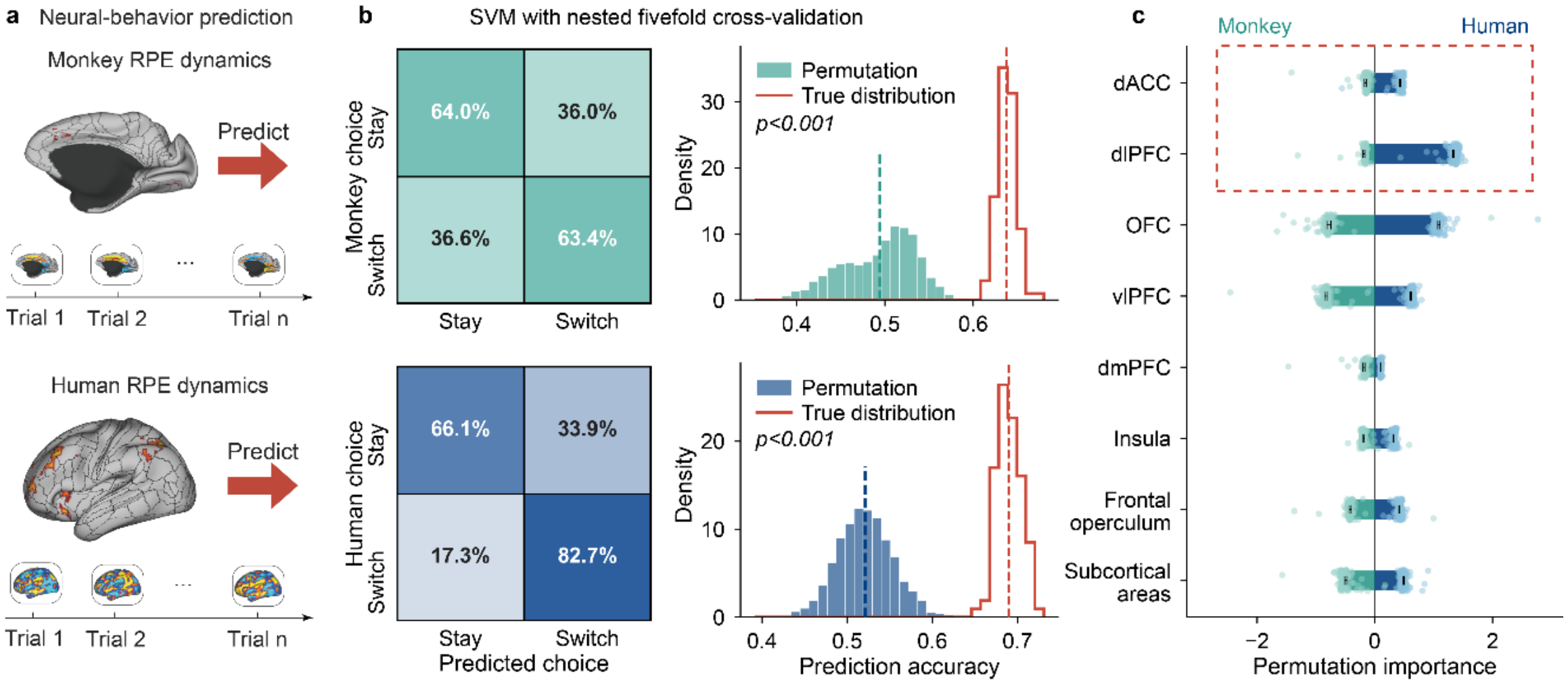

**Fig. 5: Single-trial decoding of choice behavior via dynamic RPE signals in both species.**

(a) Schematic of single-trial neural-behavioral prediction modeling. Trial-wise RPE dynamic maps (estimated by Least Squares – Separate approach) were used as input features to predict stay/switch behavior. Top: example RPE activation time series maps from macaques; bottom: analogous human data.

(b) Confusion matrices show SVM classifier performance (averaged over 100 cross-validation runs) decoding stay vs. switch decisions from RPE signals. Accuracies exceeded chance in both species (macaques: 63.8%, 95% CI 61.9–65.7%; humans: 69.0%, 95% CI 66.7–71.6%; $P < 0.001$, permutation test). Rows: true labels; columns: predictions.

(c) Regional contributions to decoding. OFC, vlPFC, dmPFC, frontal operculum, and

striatum showed conserved RPE sensitivity (contribution ratios ≈ 1), whereas dACC and dlPFC exhibited stronger human-specific involvement (dACC ratio = 2.70, dlPFC ratio = 7.30, FDR corrected $P < 0.001$) (***$P< 0.001$; **$P< 0.01$; *$P< 0.05$)

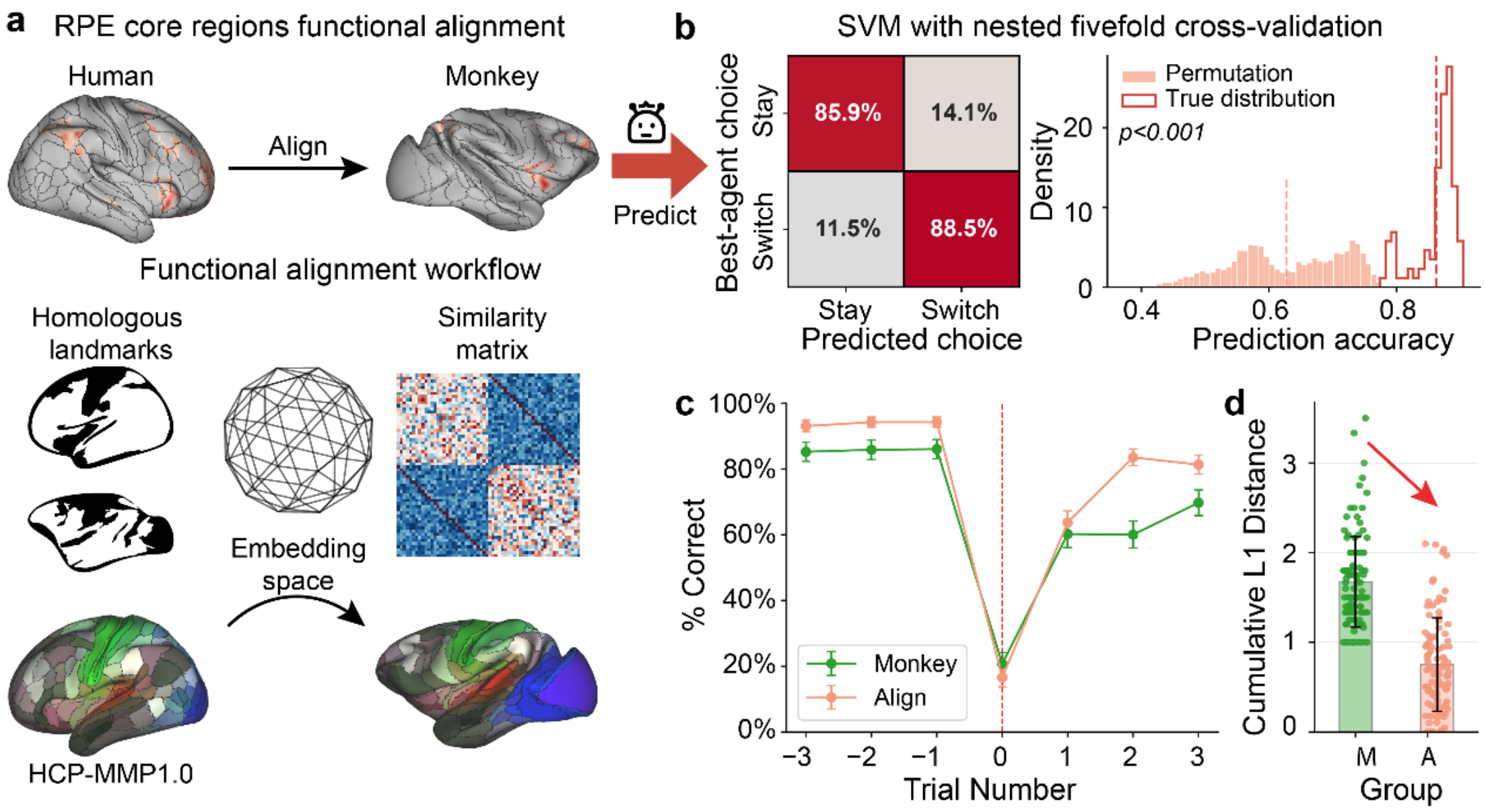

**Fig. 6: Spatial alignment to the human RPE network improves optimal behavioral prediction in macaques.**

(a) Schematic of interspecies alignment of human RPE regions to macaque brain using a joint-embedding approach (see Methods for details).

(b) SVM classifier performance predicting the best computer agent's choices from monkey RPE signals within human-aligned regions (86.3% accuracy vs. chance, $P < 0.001$; confusion matrices show mean cross-validated performance).

(c) Macaque accuracy (±3 trials around reversals): RPE-based predictions surpassed actual behavior at most lags (–1, 0, +2: $P < 0.001$; +1: n.s.).

(d) Suboptimal trials exhibited greater deviation from optimal strategies (L1 distance to

best-agent benchmark). Interspecies alignment-based predictions reduced these deviations ($t = -10.10$, $P < 0.001$), particularly near reversals.

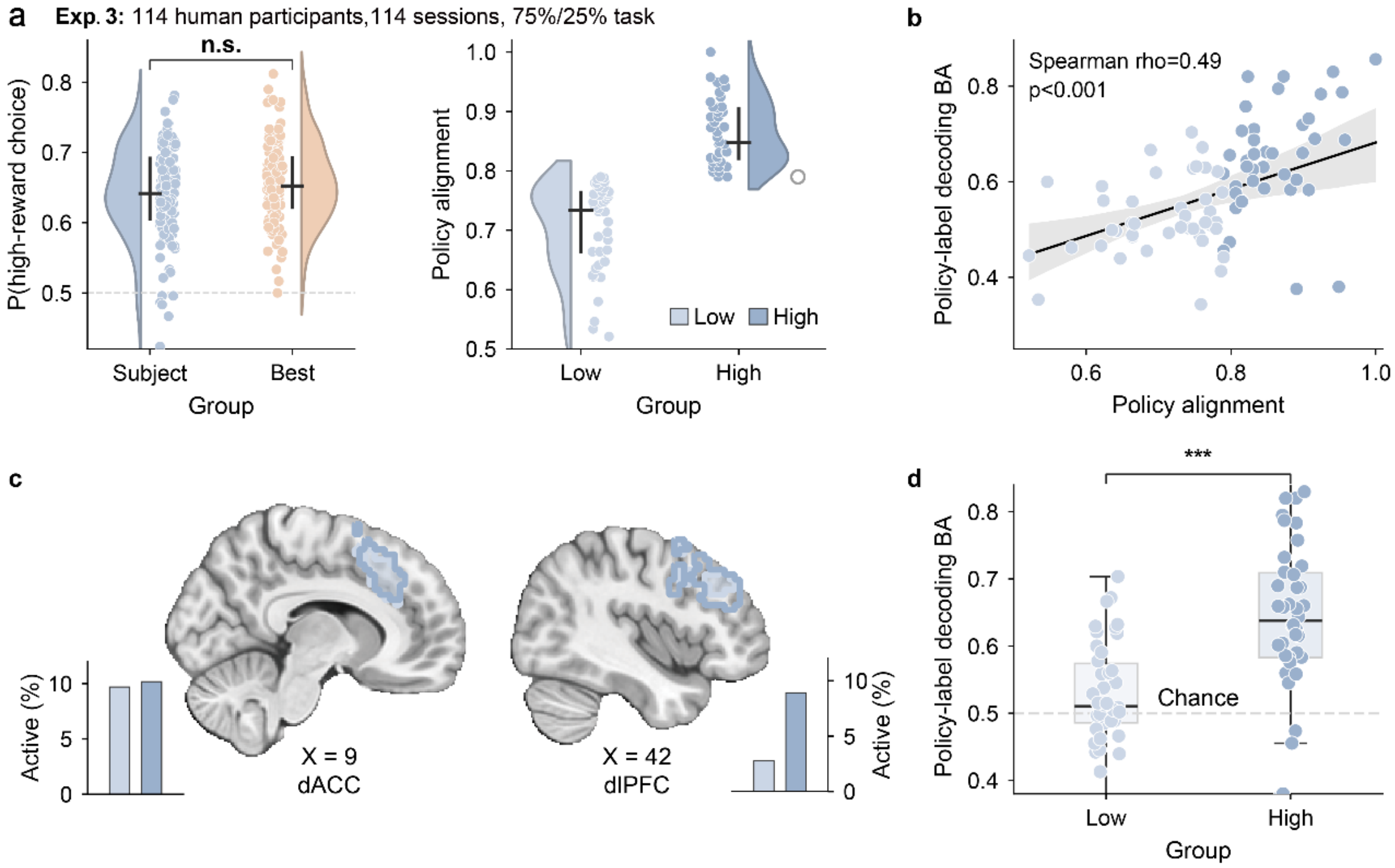

**Fig. 7: Enhanced behavioral optimization correlates with stronger prefrontal decoding of optimal policy actions.**

(a) Left: Proportion of high-reward choices across 114 human sessions and participant-specific optimal policy sequences. Bars show median ± IQR; the distributions did not differ (paired Wilcoxon W = 2,662.5, P = 0.145). Right: Optimal-policy alignment was defined from trial 3 onward as the fraction of trials in which the observed action matched the action prescribed by the k = 2 threshold policy (Fig. 2) given the preceding two trials' outcomes. After imaging quality control (85 participants), the median-ranked individual was excluded to form low- and high-alignment groups (n = 42 each; N = 84 for decoding).

(b) Balanced decoding accuracy for optimal policy actions from prefrontal RPE patterns

increased with alignment (Spearman's $\rho = 0.491$, permutation test $P = 1.00 \times 10^{-4}$). Line and shading indicate the linear fit ± 95% CI.

(c) Whole-brain FWE-corrected RPE maps ($P < 0.05$) for the low- and high-alignment groups. Proportions of significant voxels: dACC, 9.68% (low) vs. 10.17% (high); dlPFC, 2.75% (low) vs. 8.92% (high).

(d) Mean balanced accuracy: high-alignment group = 0.628 (95% CI [0.583, 0.669]); low-alignment group = 0.525 [0.501, 0.549]; group-label permutation $P = 2.00 \times 10^{-4}$. Decoding exceeded chance (0.5) in both groups (high: $P = 1.00 \times 10^{-4}$, low: $P = 0.044$; sign-flip tests). Points represent participants; dashed line marks chance level.

## Statistics

The sample sizes were similar to the sample sizes used in the field. No statistical methods were used to determine the sample size.

## Reporting summary

Further information on research design is available in the Nature Portfolio Reporting Summary linked to this article.

## Data availability

The behavioral data supporting this study are available in the zenodo repository, https://doi.org/10.5281/zenodo.17858168. The raw fMRI data supporting this study are available from the corresponding author upon reasonable request. The unthresholded statistical fMRI maps associated with this study are available in the zenodo repository, https://doi.org/10.5281/zenodo.17858168. Source data are provided with this paper.

## Code availability

The MATLAB/Python/R custom code supporting the behavioral results of this study is available in the same zenodo repository, https://doi.org/10.5281/zenodo.17858168. Any remaining code that supports the findings of this study is available from the corresponding author upon reasonable request.

## Acknowledgments

We thank Trevor Robbins, John Gore, Jian Li, Yang Zhou and Hua Tang for critical reading and constructive suggestions during the early preparation of our manuscript. We also thank Alessandro Bongioanni, Jan Grohn, Cooper D Grossman, Ido Ben-Artzi for their help during data analysis and Yilin Liu, Qian Lv, Wenwen Yu and Jie Xu for their help during data acquisition. This study was supported by Brain Science and Brain-like Intelligence Technology-National Science and Technology Major Project (No. 2021ZD0204002), the National Natural Science Foundation of China (No. U25C2016, 82151303), and grants from Peking-Tsinghua Center for Life Sciences.

## Author contributions

Z.W., F.F. and M.Y.W. contributed to the conception and design of the study. T.S., Y.C.H., F.W.Z., M.W., S.Q.Y., J.H.L., Y.Z.W., Y.J.F., M.Y.W., F.F. and Z.W. contributed to acquisition, post-processing and analysis of the data. T.S., M.W., S.Q.Y., J.H.L. and Y.C.H. analyzed all or parts of the data, and prepared Figures/tables with the help of Z.W., K.S.C., and R.S.M. Z.W. wrote the manuscript, obtained the founding and supervised the study with the help of F.F. and M.Y.W. All authors approved the final version of the manuscript.

## Conflicts of interest

The authors declare that they have no conflicts of interest.